\documentclass[nobibnotes,preprintnumbers,aps,prl,superscriptaddress,showpacs,twocolumn,twoside,sort&compress]{revtex4}				 %
\usepackage{graphicx}																												 %
\usepackage{titlesec}	
\usepackage{color}																										     %
\usepackage{fancyheadings}																											 %
\begin{document}																													 %
\preprint{{Vol.XXX (201X) ~~~~~~~~~~~~~~~~~~~~~~~~~~~~~~~~~~~~~~~~~~~~~~~~~~~~ {\it CSMAG`16}										  ~~~~~~~~~~~~~~~~~~~~~~~~~~~~~~~~~~~~~~~~~~~~~~~~~~~~~~~~~~~~ No.X~~~~}}																 %
\vspace*{-0.3cm}																													 %
\preprint{\rule{\textwidth}{0.5pt}}																											 \vspace*{0.3cm}																														 %

\title{Isothermal Entropy Change and Adiabatic Change of Temperature of the Antiferromagnetic Spin-1/2 Ising Octahedron and Dodecahedron}

\author{K. Kar\v{l}ov\'a}
\thanks{corresponding author; e-mail: katarina.karlova@student.upjs.sk}
\author{J. Stre\v{c}ka}
\affiliation{Institute of Physics, Faculty of Science, P. J \v{S}af\'arik University, Park Angelinum 9, 040 01 Ko\v{s}ice, Slovakia}
\author{T. Madaras}
\affiliation{Institute of Mathematics, Faculty of Science, P. J. \v{S}af\'arik University, Jesenn\'a 5, 040 01 Ko\v{s}ice, Slovakia}

\begin{abstract}
The spin-1/2 Ising octahedron and dodecahedron with a unique antiferromagnetic interaction display an outstanding magnetization jump at zero magnetic field, which consequently leads to a giant magnetocaloric effect during the adiabatic demagnetization. In the present work we report temperature dependences of two basic magnetocaloric response functions: the isothermal entropy change and the adiabatic change of temperature. It is shown that the Ising octahedron and dodecahedron generally exhibit a large negative isothermal entropy change upon increasing of the magnetic field, which serves in evidence of their cooling performance.

\end{abstract}


\maketitle
\section{Introduction}
The magnetocaloric effect (MCE) is most commonly characterized through the isothermal entropy change $\Delta S_{iso}$ and the adiabatic temperature change $\Delta T_{ad}$ of a magnetic solid in response to a variation of the applied magnetic field. An importance of the MCE consists in its potential usage for a magnetic refrigeration through the process of adiabatic demagnetization \cite{spichkin}. The magnetic materials with the large MCE are therefore of technological interest with regard to their cooling capability either at room or low temperatures \cite{rank04}. The large isothermal entropy change were experimentally found for diverse magnetic systems as cobalt-based chain CoV$_{2}$O$_{6}$ \cite{nand16}, gadolinium ladder \cite{chen13}, oligonuclear spin cluster Ln$_{6}$Mn$_{12}$ \cite{liuj13} or molecular nanomagnets \cite{evan09}. The isothermal entropy change is most commonly obtained either from magnetization data by means of Maxwell relation or from an integration of the specific-heat data. 

In the present work we will examine the isothermal entropy change and the adiabatic temperature change of the antiferromagnetic spin-1/2 Ising octahedron and dodecahedron, which can be calculated from the exact analytical results reported in our previous study \cite{stre15}.
\section{Model and Method}
\begin{figure}[h!]
\begin{center}

\includegraphics[width=0.23\textwidth]{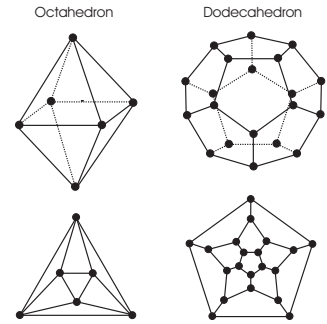}
\vspace{-0.3cm}
\caption{\small The Ising octahedron and dodecahedron. The upper panel displays a steric arrangement, while the lower panel shows
the equivalent planar lattice graphs.}
\label{fig1}

\end{center}
\end{figure}
Let us consider the spin-1/2 Ising octahedron and dodecahedron (see Fig. \ref{fig1}), which are defined trough the following Hamiltonian
\begin{eqnarray}
{\cal H}=J \sum_{\langle i,j \rangle}^{N_b} S_i S_j - h \sum_{i=1}^{N} S_i.
\label{ham}
\end{eqnarray}
Here, $S_{i} = \pm 1$ denotes the Ising spin placed at $i$th vertex of a regular polyhedron, the first summation accounts for the antiferromagnetic interaction $J>0$ between adjacent spins, the second summation accounts for the Zeeman's energy of individual magnetic moments in the external magnetic field $h>0$ and finally, $N(N_{b})$ stands for the total number of spins (bonds) of a regular polyhedron. 

The magnetization curves and thermodynamic properties of the regular Ising polyhedra defined through the Hamiltonian (\ref{ham}) were comprehensively studied in our previous works \cite{stre15, karl16}. In the present paper we will focus our attention on magnetocaloric response functions of the Ising octahedron and dodecahedron (see Fig.~\ref{fig1}), which exhibit a giant magnetocaloric effect close to zero magnetic field due to an abrupt magnetization jump towards the one-third and one-fifth plateau, respectively \cite{stre15}. The isothermal entropy change can be calculated as a difference of the entropy at non-zero and zero magnetic field $\Delta S_{iso} = S_{2}(T, h \neq 0) - S_{1}(T, h = 0)$ at the constant temperature. The molar entropy can be easily obtained from the exact results (6)-(10) presented in Ref. \cite{stre15} for the Gibbs free energy $g$ according to  
\begin{eqnarray}
S = - N_{\rm A}\frac{\partial g}{\partial T},
\label{entropy}
\end{eqnarray}
where $N_{\rm A}$ denotes Avogadro's number. The adiabatic temperature change $\Delta T_{ad} = T_{f}(S, h_{f} \neq 0) - T_{i}(S, h_{i} = 0)$ can be obtained from the transcendent equation 
\begin{eqnarray}
S(T_{i}, h_{i}=0) = S(T_{f}, h_{f} \neq 0)
\label{condition}
\end{eqnarray}                                         
which can be numerically solved by the bisection method. Here, $T_{i}$ ($T_{f}$) and $h_{i}$ ($h_{f}$) denote the initial (final) temperature and magnetic field for a given adiabatic process.

\section{Results and discussion}
\begin{figure}[t]
\begin{center}
\includegraphics[width=0.36\textwidth]{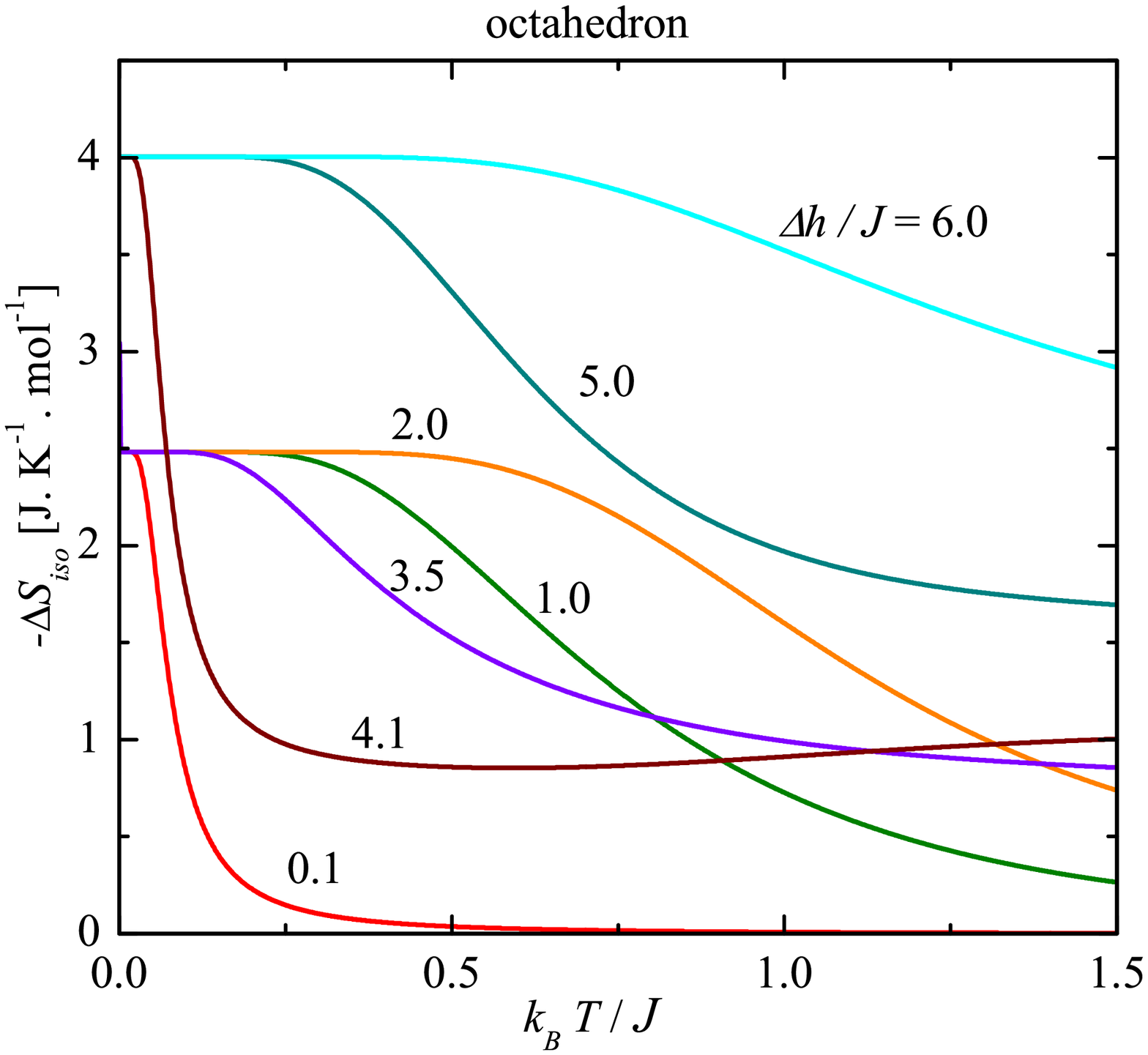}
\includegraphics[width=0.36\textwidth]{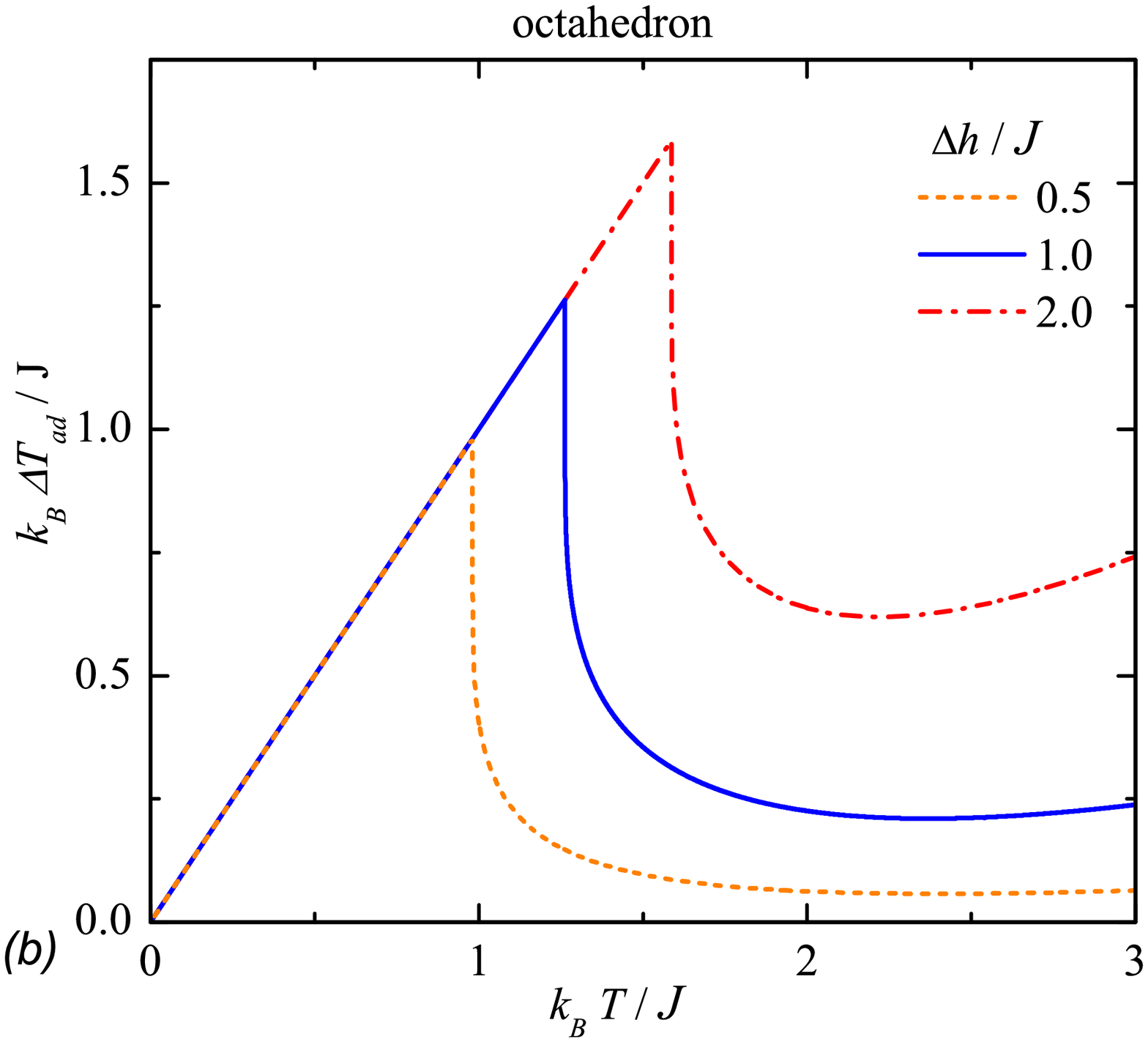}
\vspace{-0.5cm}
\caption{\small (a) Temperature dependence of the isothermal entropy change of the Ising octahedron at a few different values of the magnetic-field change; (b) temperature dependence of the adiabatic change of temperature of the Ising octahedron for three different values of the magnetic-field change.}
\label{fig2}
\end{center}
\end{figure}

Let us proceed to a discussion of temperature dependences of the isothermal entropy change and adiabatic temperature change of the antiferromagnetic spin-1/2 Ising octahedron and dodecahedron. The isothermal entropy change of the Ising octahedron is plotted in Fig.~\ref{fig2}(a) for a few different values of the magnetic-field change. As one can see, the isothermal entropy change of the Ising octahedron is negative for arbitrary magnetic-field change independently of temperature. This result implies a substantial cooling performance of the Ising octahedron in thermal contact with the heat bath. The zero-temperature asymptotic  values of the isothermal entropy change are in agreement with the difference of ground-state degeneracy at zero and non-zero magnetic fields. If the magnetic-field change is lower than the saturation field of the Ising octahedron $h_{c}/J =4$, then, one finds a gradual increase of the working temperature interval with increasing of the magnetic-field change until roughly a midpoint of the one-third magnetization plateau is reached \cite{stre15}. However, the largest isothermal entropy change of the Ising octahedron can be found for the magnetic-field change, which is higher than the saturation field. The maximum value of isothermal entropy change of the Ising octahedron for $\Delta h > h_{c}$ is $-\Delta S_{iso} \approx 4.01{\rm JK^{-1}mol^{-1}}$ at low enough and moderate temperatures ($k_{\rm B} T/J \approx 0.3$ for $\Delta h/J = 5.0$), whereas one observes its gradual decrease at higher temperatures ($-\Delta S_{iso} \approx 1.6{\rm JK^{-1}mol^{-1}}$ for $k_{\rm B} T/J \approx 1.5$ at $\Delta h/J = 5.0$).

The adiabatic temperature change of the Ising octahedron linearly increases with temperature up to the threshold temperature, which corresponds to the residual entropy $S =\frac{R}{6}\ln18$ ($R$ is universal gas constant) of the Ising octahedron for a given magnetic-field change, see Fig.~\ref{fig2}(b). Consequently, the Ising octahedron achieves the absolute zero temperature during the adiabatic demagnetization process whenever the initial temperature is selected below the threshold temperature forming a cusp in the relevant temperature dependence. This result is apparently in contrast with one of the formulations of the third law of thermodynamics \cite{grei35}.
On the other hand, there is a rapid decrease in the adiabatic temperature change above the threshold temperature, which indicates a less efficient cooling. The threshold temperature, which corresponds to the residual entropy for a given magnetic-field change, thus defines the working temperature interval of the Ising octahedron that monotonically extends with increasing of the magnetic-field change.

\begin{figure}[h!]
\begin{center}

\includegraphics[width=0.36\textwidth]{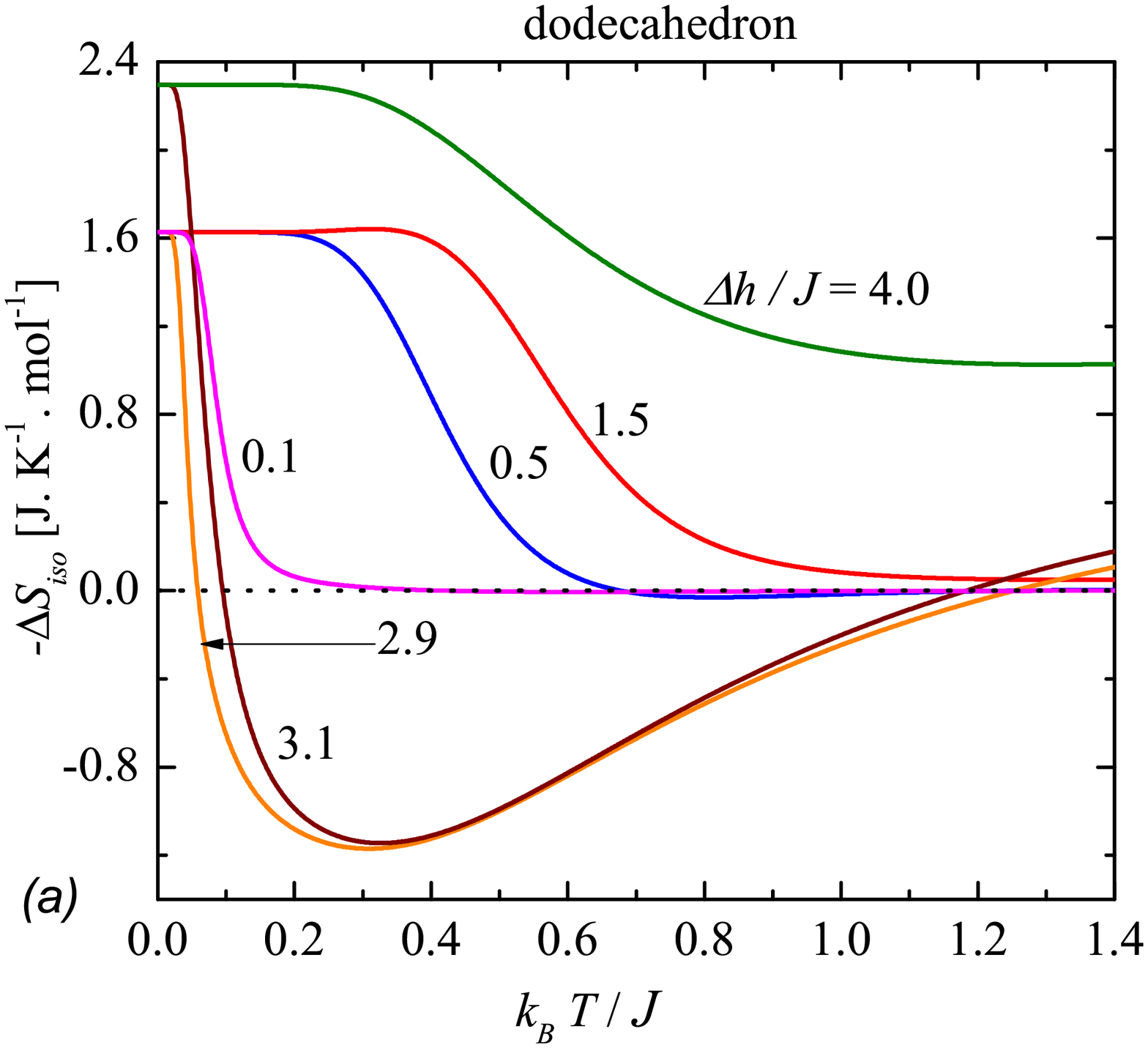}
\includegraphics[width=0.36\textwidth]{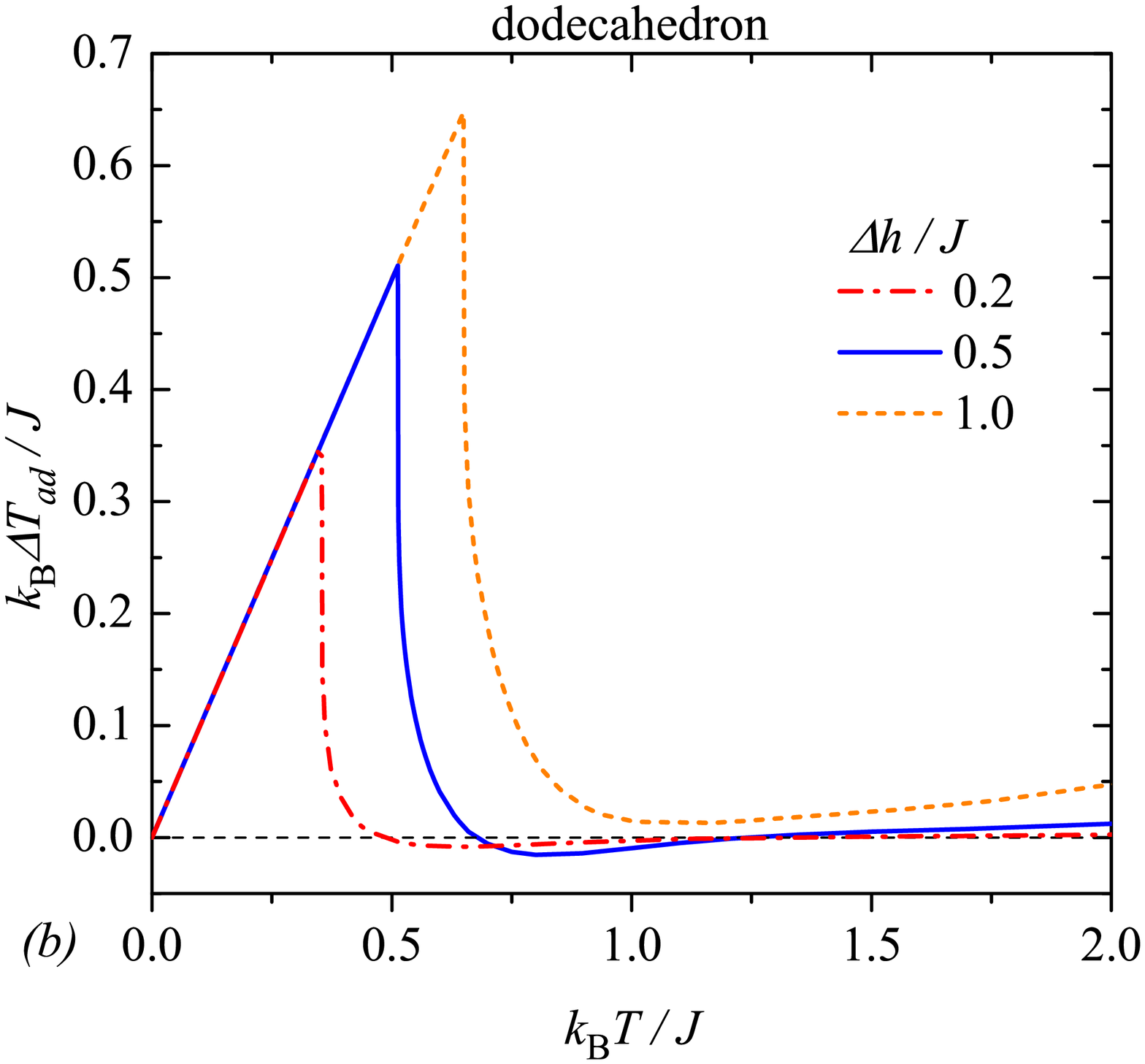}
\vspace{-0.7cm}
\caption{\small (a) Temperature dependence of the isothermal entropy change of the Ising dodecahedron at a few different values of the magnetic-field change; (b) temperature dependence of the adiabatic change of temperature of the Ising dodecahedron for three different values of the magnetic-field change.}
\label{fig3}

\end{center}
\end{figure}

The antiferromagnetic spin-1/2 Ising dodecahedron also shows large MCE $-\Delta S_{iso} > 0$ for most of the magnetic-field changes, see Fig. \ref{fig3}(a). The notable exception to this rule represent the field changes approximately equal to the saturation field $\Delta h/J \approx h_{c}/J=3$, for which the inverse MCE $-\Delta S_{iso} < 0$ can be detected in a relatively wide temperature range. The most efficient cooling through a thermal contact with the Ising dodecahedron can be thus achieved just for sufficiently high values of the magnetic-field change $\Delta h/J \geq 4.0$, because the maximal value of isothermal entropy change $-\Delta S_{iso} \approx 2.30{\rm JK^{-1}mol^{-1}}$ then persists up to moderate temperatures ($k_{\rm B} T/J \approx 0.35$ for $\Delta h/J = 4.0$). Contrary to this, the isothermal entropy change decreases down to $-\Delta S_{iso} \approx 1.0{\rm JK^{-1}mol^{-1}}$ at much higher temperature ($k_{\rm B} T/J \approx 1.3$ for $\Delta h/J = 4.0$).

The adiabatic temperature change of the Ising dodecahedron has a similar behavior as that of the Ising octahedron [see Fig.\ref{fig3}(b)]. For  arbitrary magnetic-field change there exists the threshold temperature corresponding to the residual entropy $S =\frac{R}{20}\ln250$, under which the Ising dodecahedron reaches the absolute zero temperatures during the adiabatic demagnetization. Contrary to the Ising octahedron, the adiabatic temperature change of the Ising dodecahedron may become negative at moderate temperatures as a result of the inverse MCE. In addition, the adiabatic temperature change is at the same magnetic-field changes less than a half of the adiabatic temperature change of the Ising octahedron. In this respect, the Ising octahedron is more efficient refrigerant than the Ising dodecahedron, since it provides a higher gain of the isothermal entropy change and the adiabatic temperature change at the same values of magnetic-field change due to an absence of the inverse MCE.  

\section{Conclusions}

In the present work we have exactly examined two basic magnetocaloric characteristics of the Ising octahedron and dodecahedron: the isothermal entropy change and adiabatic change of temperature. It has been demonstrated that the largest isothermal entropy change can be found whenever the field change is greater than the saturation field of the relevant Ising spin cluster. The adiabatic change of temperature confirms an outstanding cooling capability of the Ising octahedron and dodecahedron, which may reach during the adiabatic demagnetization absolute zero temperature. The Ising octahedron and dodecahedron can be therefore regarded as promising frustrated spin structures for magnetic refrigeration, whereas the maximum of the adiabatic change of temperature shifts to higher temperatures with increasing of the magnetic-field change.  

\section{Acknowledgement}
This work was financially supported by grant Nos. VEGA 1/0043/16, APVV-14-0073 and VVGS-PF-2016-72606.

\end{document}